\documentclass{sig-alternate}

\usepackage{pslatex,portland}		
\usepackage{color}		
\usepackage{url}		
\usepackage{epsfig}		
\usepackage{multirow}
\usepackage{array}
\usepackage{float}

\newcommand{\inserteps}[3]{
\begin{figure*}[htbp]
\begin{center}
{\hbox{\vbox{\makebox[\columnwidth]{\epsfbox{#1}}}}} \vspace{-0.3cm} \caption{\em #2}
\vspace{-0.2cm} 
\label{#3}
\end{center}
\end{figure*}
}

\begin{document}

\title{An Investigation into the use of Images as Password Cues }

\numberofauthors{3}
\author{
\alignauthor
Tony McBryan\\
       \affaddr{School of Computing Science}\\
       \affaddr{University of Glasgow}
\alignauthor
Karen Renaud\titlenote{Corresponding Author}\\
       \affaddr{School of Computing Science}\\
       \affaddr{University of Glasgow}\\
       \email{\tt karen.renaud@glasgow.ac.uk}
\alignauthor
J. Paul Siebert\\
       \affaddr{School of Computing Science}\\
       \affaddr{University of Glasgow}
}

\maketitle

\begin{abstract}
Computer users are generally authenticated by means of a
password. Unfortunately passwords are often forgotten and
replacement is expensive and inconvenient. 
Some people write their passwords down but these records can easily be lost 
or stolen.
 The  option we explore  is to find a
way to {\em cue }
 passwords securely. The specific cueing technique we report on
 in this paper 
employs images as cues. The idea is to elicit textual descriptions
of the images,
which can then be used as 
passwords. 

We have defined a set of metrics for the kind of image that could
function effectively as a password cue. We  identified five candidate image
types and ran an experiment to identify the image class with the best
performance in terms of the defined metrics. 

The first experiment identified inkblot-type images as being superior.
We tested this image,
 called a {\em cueblot}, in a real-life environment.
 We allowed users to tailor their cueblot until they
felt they could describe it, and they then entered a description of
the cueblot as their password. The cueblot was displayed at
each subsequent authentication attempt to ``cue'' the password.
Unfortunately, we found that users did not exploit the cueing potential
 of the 
cueblot, and while there were a few differences between textual
 descriptions of cueblots
 and non-cued passwords, they were not compelling.

Hence our attempts to alleviate the difficulties people experience
with passwords, by giving them access to a tailored cue, did not have
the desired effect. We have to conclude that the password mechanism
might well be unable to benefit from  bolstering activities such as
this one.

\end{abstract}

\section{Introduction}

Computer users of the 21st century cannot escape
the need to authenticate themselves.
 This is mostly
achieved by means of a secret password. Since people use a multitude
of systems, 
 people have to remember passwords for each of these. 
Since human memory is fallible, the direct consequence of this is that
people forget their passwords, and need to have reminders or
replacements. 

In this paper we address the issue of password cueing. This term may seem
to be an
oxymoron since passwords are a security tool, 
intended to protect some data or service, and need to remain secret at
all times. Cues,
if at all clear and  helpful, would tear a large hole in the security
ostensibly 
maintained by the password. 

We do, however, believe that cues could be provided in the form of an
abstract image so that the cue itself is so obscure and vague that it
acts as a cue only to the legitimate owner of the password. The cue
acts only as a cue and not as a hint which could lead a potential imposter
to the password. Obviously the kind of image used is critical, and we
have therefore conducted an experiment to test different kinds of
images to determine which is the most viable. We then used the
resulting images in a real-life setting in order to determine their
efficacy and usefulness to the end-user. 

In Section \ref{auth} secret-based authentication  is reviewed.
In Section \ref{forget} we provide a brief overview of the literature
related to forgetting 
and cueing.
In Section \ref{vision}, we motivate the use of images as cues and explore
the
kinds of images that could be used in terms of memorability and
diversity of text associations. 
Section \ref{imagcueing} provides a synopsis of the previous sections
and formulates the research question. 
Section \ref{method} gives information about the methodology followed
in order to find the best cueing image.
Section \ref{results} presents the results of the experiment and
Section \ref{discuss} presents the results and identifies the best
image type.
 Section \ref{passwords} 
reports on the  experiment that tested the use of the best image type as
a cue during authentication. 
Section \ref{conc} concludes.

\section{Authentication}\label{auth}
In order to grant access to a restricted digital space, we use a two
phase protocol: identification followed by authentication. Users are
identified by means of a text string --- either an email address or a
special user name --- and then authenticated to verify the
identity. During authentication the user's identity can be verified by means of a shared secret, called a
key, or by means of a biometric which measures the
user's physiology or behaviour and matches it to a previously recorded
template. Since biometric measuring devices are more expensive than
keyboards most authentication these days is done by means of a
shared secret password. 

There is nothing inherently wrong with this  --- but in the
face of fallible human memory and insecure communication channels it
tends to fail. The undeniable fact is that 
people often forget their keys and these have to be replaced \cite{Witty04}. 
Unfortunately, there are some
problems with current replacement practices:
\begin{enumerate}
\item 
The
replacement process weakens the mechanism because a replacement key has to be
delivered in some way and this delivery can be intercepted by an
intruder who then proceeds to impersonate the legitimate user. If
challenge questions are used the mechanism is weakened unacceptably,
as we shall discuss in the following section. 
\item The replacement has to be funded and the cost is anything but
  negligible. Gartner \cite{Witty04} claims that a single replacement
  costs between \$15 and \$30. They estimate that each employee will
  call about 5 times a year (since they have passwords for multiple
  systems).  A cheap alternative is simply to send people their
  passwords by email, but since email is seldom encrypted, this option
  can only be used for insecure systems, and only where people haven't
  forgotten their email password. 
\end{enumerate}
\noindent What we
need to do, therefore, is to devise a way to help users to remember their
passwords. The traditional way is by means of a physical record but
this is extremely insecure and should be discouraged. 


Since forgetting is the ``fly in the ointment'', the next section
takes a closer look at this human propensity. 

\section{Forgetting}\label{forget}

Humans learn in two ways --- explicitly and implicitly. Implicitly
learnt skills seldom decay but explicitly learnt knowledge, the
category passwords belong to, is
extremely prone to decay.
Unfortunately humans  {\em do} forget their passwords, and the
forgetting is seldom deliberate.  
Ebbinghaus \cite{Ebbing} proposed a {\em forgetting} curve, which
predicts that most forgetting occurs early on in the process and then
slows down later on. Thus details may be forgotten within minutes if
no serious attempt is made to encode the information in more than a
cursory fashion. 

Consider, now, how most passwords are chosen. Someone
visits a website and is asked to provide a password, which is to be
used at future visits for authentication purposes. The person's goal
is to peruse the website, or purchase some items, or perhaps something
else --- but whatever it is, the definition of the password is
probably extraneous to the person's immediate goals and purposes. If the person
has had experiences of forgotten passwords in the past, a 
 well-worn password that is used for this kind of eventuality might be
 provided.
 If there is a concern about security and the user is wary of
using a previously-used password, she/he may provide a unique password
and write it down. If, however, no
 record is made, the
password is likely to  be forgotten, especially if
 the site is used infrequently. Since people tend to rely on their
memory to retain passwords \cite{Gaw06}, this is  the most likely scenario.

Schacter \cite{Schacter01} calls this forgetting the {\em sin of
  transience}. Researchers and practitioners have tried to come up with ways of improving
  memory. Schacter cites a number of memory improvement programs and
  health cures and concludes that none are miracle cures.
  One thing that {\em does} assist effective retrieval of remembered
  facts  is the effectiveness of the encoding process.
 Schacter cites research into a mechanism called {\em
  elaborative encoding} where the person spends some time encoding the
  information using visual imagery, mnemonics
   or elaborative questions. These are indeed effective but, of course,
  require extra effort
  and are unlikely to be used in an uncontrolled password defining
  setting. Experience shows that password users seldom take this
  trouble \cite{Gaw06}.

Another way of improving retrieval is by the provision of cues.
Nyberg {\em et al.} \cite{Nyberg00} 
argues that retrieval of information activates the same brain regions
as those activated when the information was encoded. They experimented
with word-sound encoding and found that provision of the sounds
assisted word retrieval. Moscovitch and Craik \cite{Mosco76} found
that cueing was beneficial at deeper levels of encoding.  
The following section considers cueing mechanisms with emphasis on
their use in an authentication context.

\subsection{Cueing Mechanisms}\label{cues}
A cue can be defined as:
\begin{quote}
a. A reminder or prompting, or\\
b. A hint or suggestion.
\end{quote}

A cue heard by someone other than the person for whom it is intended,
therefore, could produce the same association or act as the same
reminder as it was intended to elicit in the target person ---
especially if the cue is effective. In an
authentication setting such a universal cue is useless since it undermines
the security of the authentication key. Thus a cue used in an
authentication setting needs to
be deliberately obtuse. It should make sense only to the legitimate
user, and not to anyone else.

One of the most common mechanisms, used by a variety of websites, is
that of challenge questions. On the face of it this is a viable
mechanism for proving identity when passwords are forgotten. A closer
look at challenge questions reveals many flaws. One has two choices in posing questions --- either the user
  chooses his or her own
questions at enrollment, and provides the answers, or the system has a
  set of questions, and the user is allowed to choose one, and 
  provide the answer. Both options have problems: 
\begin{itemize}
\item If the user has to generate the question he or she/he is equally
  likely to forget the question as the password. In this case the cue
  question places an extra demand on the user's memory and is equally
  vulnerable to decay. One also has to put software in place to
  ensure that users specify reasonable and well-formed questions and
  do not simply enter their own password as the
  question, for example.
\item If the system has a set group of questions these need to be
   applicable to a wide range of users. Thus the site owners resort to
   setting widely applicable questions based on, for example, the name
   of the person's first school, first pet or
   mother's maiden name. The fatal flaw with these questions is that a
   relatively superficial knowledge of the legitimate user is required
   in order to know the answers to these questions, and the challenge
   questions thus offer an intruder a convenient and insecure way into
   the system.  Even if the answer to the chosen question is not
   easily determined using research, the fact  that
   most sites revert to the same set of questions reduces the
   ``secrecy'' of the answer. The more  sites
   holding the answers the less secret they are. 
\end{itemize}
\noindent Other sites prefer not to make use of challenge questions and revert
to emailing forgotten passwords to users. This is also an insecure
practice because email is seldom encrypted and is easily intercepted
by a hacker. The use of one-time passwords, which
require changing as soon as the person logs in, is somewhat more
secure, but only if it is indeed the legitimate user who is trying to gain
access. If an intruder is requesting the password reminder, and
watching for the email, the legitimate user will probably be
completely unaware of the intrusion into his account until the
negative effects of the intrusion manifest themselves.

Since email reminders and challenge questions are insecure and ineffective we should attempt to find some other
mechanism of reminding users of their forgotten passwords. One way of
doing this is by providing the user with a cue, as has been done by
Hertzum \cite{Hertzum06}. He proposes that users specify particular
password characters which will be displayed at password entry in order to
jog their memory. This idea was tested with 14 users and it was found
that it did help them to remember their passwords. Hertzum notes, however, that
the defined passwords were often weak and predictable and argues that
some kind of cueing mechanism is required in order to support the use
of longer and stronger passwords. 

The possibility we have explored  is the use of images as cues. 
There is strong evidence that pictures are  more memorable than words. 
This
commonly known  {\em picture superiority effect}
claims that images are stored
with names, or labels, associated with them
\cite{Paivio}, which enhances memorability.
The
idea for our research is that the user 
is given a personal image at enrollment and that the password
would essentially be the {\em image description} or {\em name}. The user
can then request 
the image to display as a password reminder should she/he forget the
password. 
A purely
representational image will not work in this secure context because what one
really needs is an image that elicits a different textual
association from different users so that intruders cannot confidently
guess textual 
associations within the three strikes allowed before a lockout. 

Stubblefield and Simon \cite{Stubblefield04} experimented with
using inkblots to assist users to form a semantic
association with the textual password, which could be used as a reminder
mechanism as required. They displayed 10 inkblots in a
particular sequence. For each blot the user was required to enter two
characters --- the starting and ending character of their inkblot
description. They had some success in trials of this mechanism,
achieving an entropy of 4.09 bits per character. However, the
cognitive load imposed on the user is significant. They do not merely
provide a textual description; they have to parse it in their minds to
extract the required starting and ending character, and then type that
in. Stubblefield and Simon do not give demographic information about
their experimental subjects but one can envisage this cognitive load
being untenable for any but the most mentally agile of users. 

There is evidence, however, that passwords based on associative memory
are more memorable but harder for people to guess
\cite{Podd96}. In addition to Stubblefield and Simon's proposal
outlined above, associative passwords have been trialled
for sound 
clips \cite{Liddell03} and for other words \cite{Smith87,Podd96}. Sound
associations were not particularly successful 
they tested the association between sound and an image. The system was
evaluated by a group of students in a lab, which made the sound
problematical. In the end the students simply memorised the pictures
and did not use the sound to cue the pictures. 
Word association works well, but is very time consuming, both
at enrolment and authentication. 

Our hypothesis is that we could make use of  images as direct cues,
without the intermediate processing required by Stubblefield and
Simon. We therefore need to determine what kind of image could support
this cueing activity in an authentication setting.
We need to find out what characteristics  this image would have to exhibit to
  facilitate superior recall in the authentication context.
The image descriptions would also have to be more durable than random
textual passwords in order to improve the current situation.

Von Ahn and Dabbish \cite{Ahn04} did research which relied on similar
skills, but for a different purpose. They constructed a game, which
required people to label images, all the while attempting to guess at
the labels  other game players have used to describe the
image. Since their main purpose is to find commonly used descriptive labels for
images, the research is different from ours. Our main purpose is to
identify an image type which will elicit very different descriptions
from different people. 

 We conducted a series of experiments 
in order empirically to verify the use of
images in this context.
Before discussing our experiments, we need first to
discuss different image types and  the
effects of human vision on the image choice.

\section{Human Vision}\label{vision}
One of the most vital of the human senses is vision. 
When  an object is seen, the viewer will compare that
object to an internal ``database'' of objects within his or her mind, and use
past experience to match that object with the object being seen in order
to identify it. Thus visual perception interacts with perceptual
processes but also with memory,
reasoning and communication \cite{Jacob03}. 


This research considers the use of images as cues. In order to act as a cue in 
an authentication environment, the image must have the following characteristics:
\begin{enumerate}
\item {\em Ambiguity} --- The image cue should mean different things to different people.
Thus a straightforward representational image is unsuitable if the users share a common language since
the description of the image is likely to be similar. Hence we need some kind
of ambiguous image to act as a cue. 
An ambiguous image is interpreted differently by different people, according to the individual's
particular perceptual processes and past experiences of the world. 


Hence if we can identify this kind of image, a specific user's cue will not necessarily be useful
to an intruder.

\item {\em Efficacy} --- Human memory for pictures and their textual
  description needs to be superior to word memory so
  that the cueing mechanism is meaningful and excites  a durable
  association. Furthermore, the textual description needs to be
  strong enough to act as a password. 
\end{enumerate}

The following two sections address these concepts in greater detail. 

\subsection{Ambiguity}

\label{gestalt} A group of
psychologists called the Gestalt psychologists have formulated a set
of laws of organisation that help us understand the  perceptual filling-in
process. 
 The laws relate to
\cite{Bruce90}:
 {\em Closure}, {\em Good Continuation}, 
 {\em Proximity}, 
 {\em Similarity}, 
 {\em Relative Size, Surroundedness, Orientation and Symmetry} and
 {\em Common Fate}.
To achieve ambiguity in the authentication context we need to find
images that are sufficiently vague in terms of the 
Gestalt laws so that they will lead to multiple interpretations.

There is a category of images called ``Ambiguous Images''. Bruce and
Green \cite{Bruce90} give some examples of pictures that depict
different things depending on foreground/background ambiguity. They
point out that humans see either the one or the other, but not both at
the same time. These pictures have deliberately been made ambiguous and do
not serve our purposes very well since they usually have only two possible
interpretations. The first question we need to ask, in choosing images for
this research, is how humans make sense of images that are not
obviously representations of a particular object.

We therefore
need a way to describe different candidate image types so that we can arrive at a particular
description of an efficacious  image type that can act as a cue.
Alario and Ferrand 
\cite{Alario99} have classified a number of images and propose the
following norms to describe them:
\begin{itemize}
\item {\em Name agreement} --- the degree to which the people agree on the
  name of the picture;
\item {\em Image agreement} --- the degree to which the person's mental image
  matches the picture;
\item {\em Familiarity} --- the familiarity of the concept being depicted;
\item {\em Visual complexity} --- measuring the number of lines and details in
  the picture; 
\item {\em Image variability} --- indication of whether the name of an object
  invokes many or few images for the object.
\end{itemize}

\noindent These norms will be used in later sections to delineate the kinds of
images the cueing application requires.  
Broadly speaking, we can use them to identify
the best image genre for cueing purposes. Obviously representative
images will have high name agreement and this disqualifies them. We
are therefore left with the broad class of {\em abstract images} ---
images not directly representing a real object. Within
this class of 
images we need to identify particular classes that are sufficiently
vague in terms 
of the Gestalt laws to make them abstract but not so vague that they
are impossible to name. If we are able
to identify such an image class, our next concern is the efficacy of
the textual description a particular image member of that class will
elicit, in terms of acting as a password cue. 

In addition to abstract images, we decided to include a special image
  type ie. {\em human faces}, in our experiment. The face is a special image as
  far as humans are concerned. Each face has the same configuration
  and elements and yet humans are able to identify thousands of faces
  without any difficulty. There is disagreement amongst researchers
  about whether faces are processed as a unit or in terms of component
  features \cite{Ellis81}. Smith and Nielsen \cite{Smith70} argue for
  a two-phase recognition process  --- a holistic processing phase
  followed by a process which does a feature by feature
  comparison. Dodson {\em et al.} \cite{Dodson97} found that when people
  were required to 
  come up with a description of facial features it impaired their
  ability to recognise the face at a later time. It seems reasonable
  to conclude that forcing people to
  describe individual components or features of faces is
  detrimental to the memory encoding process. On the other hand, Bower and
  Carlin \cite{Bower74} found that if people were asked to attribute
  intelligence to different faces, they remembered the face better
  later on. This is perhaps because the process of attributing
  semantic codes to the faces requires additional processing and this
  helps to encode the face in the person's memory whereas the previous
  study considered features in isolation and ``whole face'' encoding was not
  encouraged or facilitated. We have included
  faces in our study to see whether the textual descriptions people
  ascribe to faces meet our requirements.

\subsection{Efficacy}
In order to determine viability of a particular image class as a cueing
mechanism,  we need  a way to judge the efficacy of
textual descriptions of members of the different image classes.
This efficacy encompasses more than one aspect:

\begin{enumerate}

\item {\em Descriptiveness} --- Humans should have the ability to describe the pictures in a
  textual format --- this is termed {\em picture naming}.
\item {\em Strength} --- The text association, in order to qualify as
  a strong password, needs to have either length or complexity, which
  make it harder to break. 
\item {\em Memorability \& Durability} --- Human memory for pictures
  needs to be superior to word memory so 
  that the cueing mechanism is meaningful. However, even more
  importantly, the text
  association should be durable in the sense that users 
  are able to reproduce it perfectly after a time lapse.

\end{enumerate}
\noindent We now present a summary of the literature in each of these areas.

\subsubsection{Descriptiveness}\label{descript}
Humans communicate by naming objects, a skill that is as effortless as it
is essential to speech. 
The central premise of this research has been that we can rely on the
previously-mentioned picture superiority effect {\em accompanied by}
reliably retained textual descriptions.
%
 Levelt {\em et al.} \cite{Levelt98} present a processing model of the
picture naming task, which includes the following steps:
\begin{enumerate}
\item Recognition of the visual object.
\item The person now searches
  through his or her internal  memory structures to find a match for the object.
\item During the following stage a selection is made from the internal
  structures.
\item Next there must be another matching process --- where the internal
  structure is matched to a  word representation.
  \item Now, what Levelt calls {\em syllabic gestural scores} are
  derived. This converts the chosen 
  word's phonological shape into syllables that can be articulated.
\item Articulation can only occur once all the previous stages have
  completed. 
\item Self-monitoring. Speakers can determine, during this stage,
  whether there has been an error, and self-repair. 
\end{enumerate} 

The  textual description attribution process requires
the user to enter the description via the
keyboard,
 so that the 
last two stages of the picture naming process
given above will be replaced by processes attuned to writing and not
to speaking. The self-monitoring stage is inappropriate in this setting
since passwords are not echoed to the user due to security
constraints. Bonin {\em et al.} \cite{Bonin02} state that researchers
are not agreed as to whether the 
phonological stage 
is involved in the production of writing \cite{Geschwind69,Luria70}
 or whether the orthographical codes can
be accessed directly \cite{Miceli97,Rapp97} (both sets of researchers
referenced by Bonin {\em et al.}). Bonin's research has confirmed that speaking and
writing share processing levels but that each also has a relative
degree of autonomy.

For the purposes of this discussion we can probably ignore these
differences between spoken and written picture naming. What {\em is}
important, in the context of cueing by means of  abstract image, is that
the above process will be augmented since the abstract image is more
expressive than a representative image, and does not have a
simple label, but requires the person to use specific
perceptual and cognitive processes in order to interpret, identify and
verbalise what he or 
she sees in the image and to produce  a textual
description. For example, consider the process involved in assigning a
name to one of the most famous images: inkblots. Rapaport \cite{Rapaport46}, referring to the 
Rorschach inkblot verbalisation process, argues that such a
process is an {\em ``association process initiated by the inkblots as
  stimuli''} (p91). The results of the association process need to be
converted to language, and this process is highly dependent on
individual factors \cite{Gold87}. Hence even if two people perceive a
particular 
image as belonging to the same semantic class they are likely to verbalise it
in slightly different ways. We hope that these individual differences will
lead to syntactically different picture descriptions and therefore distinctly
different passwords.

\subsubsection{Strength}\label{security}
In order to use an image as a cue, we need to consider the security aspects of
  the image. 
Passwords are generally broken in one of two ways if there is no cue: brute strength
or dictionary attack. The former simply works its way through all
possible permutations until it finds a match. The latter exploits the
fact that most people will use a recognisable word in their own
language and works its way through dictionaries until a match is
found.   
The latter approach is by far the most popular because passwords can
often be broken within a matter of minutes using this technique
whereas brute force is extremely time consuming. For example, in 2006 some hackers managed to get hold of a
  number of MySpace passwords. Security expert Bruce Schneier
  \cite{MySpace} analysed the passwords and found that the top three
  most used
  passwords were {\em password1, abc123} and {\em myspace1}.

Hence, to make it
harder for a dictionary attack to succeed we need to make the password
less susceptible to this kind of attack. There are two ways of doing
this --- either by making the password longer by using more than one
word or by making it more
complex by including numeric and other special characters. 

The latter
approach has severe memorability limitations and the technique of replacing of
vowels with numbers, such as using a 3 instead of an e, is so well
known by attackers as to be almost useless. 
Recent studies have found that it is easier for observers to gain
knowledge of this kind of password because it is harder to type in
than if the user is typing in a familiar word \cite{Tari06}.
Making the password longer, then,
appears to be the most beneficial approach.  

Since we're asking people to describe non-representational images, we would expect to see
longer passwords, which will contribute towards strength.
Furthermore, there is evidence that previously seen pictures are named
faster than 
new pictures \cite{Mitchell88}. 
Hence by timing responses a system may be able
to infer that a possible intrusion attempt is underway. 
Since abstract images may well initiate the same semantic association
in the legitimate user and the intruder, but a slightly different
syntactical conversion is produced, the best way to prevent an
intruder from trying different possible descriptions until he or she
succeeds is by judicious use of the ``three tries lockout'' policy.

\subsubsection{Memorability \& Durability}

The picture superiority effect 
states that humans remember pictures 
better and for longer than words. Psychologists have demonstrated this
with a number of experiments \cite{Madigan83,Paivio71,Paivio68}. 
Research on advertising has  shown
that recall of pictures is high even after
as little as a 10 or 30 second exposure. Singh {\em et al.}
\cite{Singh88} found that after 6 weeks people retained almost half
the products 
and a third of the claims after only two 30 second exposures.
To explain this effect Paivio proposes a {\em dual coding}
theory. This theory proposes that humans remember both a visual and
verbal code for images, and that this eases retrieval since there are
two pathways available and each provides a pathway to the
other. Nelson \cite{Nelson77} suggests that pictures 
are described in a richer more detailed fashion in memory and it is
this that leads to
 superior retention. 
Whatever the reason, there is a solid body of evidence that humans,
having seen an image once, will readily be able to attest to the fact,
and that this effect is stronger than word-related memory effects.

That said, it must be borne in mind that all these experiments have
tested {\em recognition memory} whereas the use of cueing images requires the
use of {\em recall memory}. Recognition relies on the person
identifying a previously-seen 
picture, usually from a group of pictures. Recall requires the person
to re-generate the name of a previously-seen picture. 
%
%
There is some evidence, however, that people recall picture names for a long time. Cave
\cite{Cave97} found that a single exposure to  a picture could be
detected even after 48 weeks by examining naming response times at
subsequent exposure to the image. An
interesting effect was observed by MacLeod \cite{MacLeod88}, who
studied the re-learning effect. He tested the memory of pictures in
terms of recall (where subjects had to recall the name of the picture)
and in terms of recognition (where subjects had to identify the
previously-seen picture). He determined that there
is a savings effect for pictures when a recall acquisition process was used, but not
when a recognition acquisition process was used. The savings effect is
an effect whereby people are unaware that the knowledge is available
to them until they try to relearn something. The vastly shortened
acquisition time is a result of the previous learning.

This research does not rely on recognition, but rather on recall, and the
savings effect should therefore be active. 
The use of images in this study requires the user to study the image
and to describe it --- a fairly cognitively intensive process. Our
expectation is that the details will be recalled even after a time
lapse.

\subsection{Summary}\label{imagcueing}
We have enumerated two characteristics images need to exhibit
in order to use them for cueing: ambiguity and
efficacy. In order to satisfy the first requirement we tested a
number of abstract image classes, classes of images that elicit no immediate association
with any real life object, and the face class, which has proven
memorability. We tested  a number of images from
each of of these image classes in order to determine
efficacy of the class, by analysing and testing the following:
\begin{enumerate}
\item {\em Descriptiveness}  --- to what extent is it possible for people to
  assign a name to the image? This will, to some extent, be assisted
  by the adherence of the abstract image to the Gestalt laws. In terms
  of Alario and Ferrand's norms, we need {\em high visual complexity}.
\item {\em Strength}  ---
measured in terms of
length of the description, the character distribution of the
responses, and the entropy of the description. 
We also need to test for {\em low  name agreement} and {\em high image
  variability}, which tests whether 
different people provide the same names 
  for the image or whether descriptions differ.
\item {\em Memorability \& Durability} --- How durable are the image text
associations? 
In order to determine this we will conduct an experiment to test the
 memorability of image descriptions.
Memorability is directly related to {\em high  image agreement} --- a stronger
  single mental image  will lead to higher likelihood of the user remembering
  the image description than many mental images for the same abstract image..

\end{enumerate}

 We investigated this by means of an experiment which
compared the different abstract image types in terms of convergence of
image descriptions (to measure descriptiveness and strength).
The experiment is described in the following section. We tested
the durability of the textual descriptions of the best performing
image in a further experiment, which is described in Section
 \ref{passwords}.

\section{Testing Different Image Types}\label{method}

The most suitable images for testing, which meet the requirements laid
down in Section \ref{vision}, are those that exhibit the required level
of vagueness in terms of the Gestalt
laws \cite{wertheimer:usersnotenemy} discussed in Section
\ref{gestalt}, on the one hand. We will also be using faces because of
their proven memorability.

As
explained in Section \ref{imagcueing}, we require
images that have {\em low name agreement, high image agreement}, are
{\em visually
complex} and those for which it is possible to come up with a memorable textual
description. Our Images are shown in 
Appendix \ref{images}. The relationship between the image classes and
the Gestalt laws
is shown in Table \ref{Gestalt}. 
\begin{itemize}
\item {\em  Faces} ---  Humans are famously good at remembering
faces. The reasons for this are debated by learned researchers. Some
believe that the human brain has a special ability to recognise faces \cite{Kanwisher97}
but others believe that it is a skill we are good at because we
spend a great deal of our lives doing it \cite{Gold04}. Whatever the
reason, the fact remains that humans are good at recognising
previously-seen faces. Whereas memorability is clearly not an issue,
durability might well be. Chance and Goldstein \cite{Chance76}
conducted an experiment to determine whether previously assigned
verbal labels would be recalled after a time lapse. They found
performance in recalling verbal labels to be weak and unreliable with
only 35\% of verbal labels being recalled correctly. However, despite this
we included faces to see whether our experience replicates theirs.
\item {\em  Fractals} --- Singh \cite{Singh05} quotes
   Works as saying
  that fractals are appealing to humans due to their innate
  aestheticism. This  makes the fractal a good candidate for our
  research but their suitability for cueing remains to be seen. 
\item {\em  Inkblots} --- Stubblefield and Simon
\cite{Stubblefield04} used inkblots, and gained good preliminary
results. The most famous user of inkblots was Rorschach
\cite{Rorschach}. He thought that the responses to his inkblots could
be used to assess personality. This particular test is no longer
given much credibility \cite{Wood99} but the technique for eliciting variable responses could
work very well in our context. 
\item {\em Snowflakes} --- Snowflakes were used by Goldstein and
Chance \cite{goldstein:visualmemory}
as part of a larger experiment measuring recognition ability but no work has been performed to
study users' descriptions of these images.
\item{\em Textures} --- The Texture image type was chosen because of
their intrinsic variety.
  Textures can have smooth or rough, coarse
or fine as well as having regular or irregular patterns.
\end{itemize}

\begin{table*}
\begin{center}
\begin{tabular}[h]{||p{2cm}|p{2cm}|p{2cm}|p{2cm}|p{2cm}|p{2cm}||} 
\hline
\hline
Image Type		&Clo\-sure	&Con\-tinuity		&Prox\-imity		&Simi\-larity		&Sym\-metry\\
\hline
\hline
Faces			&$\surd$	&$\surd$		&
&			&$\surd$\\
\hline
Fractals		&		&$\surd$		&
&$\surd$		&\\
\hline
Inkblot		&$\surd$	&$\surd$		&$\surd$
&$\surd$		&$\surd$\\
\hline
Snow\-flakes		&$\surd$	&$\surd$
&$\surd$		&$\surd$		&$\surd$\\
\hline
Textures		&$\surd$	&
&$\surd$		&$\surd$		&$\surd$\\
\hline
\hline
\end{tabular}
\end{center}
\caption{Image Classes \& the Gestalt Laws}
\label{Gestalt}
\end{table*}

We obtained our experimental images from multiple sources.  Our human
faces were chosen from the Essex Face
Database\cite{essex:facedatabase} to obtain an equal balance of gender
in good lighting conditions on plain backgrounds.  A commercial
toolset\cite{slijkerman:ultrafractal} was used to generate the fractal
images.  Inkblots were generated using a custom parameterised script
we developed to control the distribution of blots within the image.
Snowflakes were generated using a free Snowflake generator
tool\cite{aistudio:snowflakegen}.  Textures were chosen from the
CuReT \cite{curet} texture database to comprise of both artificial and
natural objects with varying degrees of lighting conditions and
smoothness of the texture.  Generated images were chosen by varying
parameters which resulted in images which were distinctly different in
appearance. 

We developed
 a web-based application to present participants with the images shown
 in Appendix \ref{images} so
that they could provide a textual association for each.  
We allowed
users the option of not 
submitting an association for a particular image if they found it
difficult to form an association.
We 
 collected all 
associations and performed an analysis, which is reported in the
 following Section.



\section{Results}\label{results}
We first consider the response rate for images in terms of the number
of responses gathered for particular images. The
user could choose not to provide an association for an image and this
act of not responding to an image was taken as an indication that the
user found the image too difficult to generate a textual association
for.  As such, this serves as an implicit subjective measure related to the
ease of forming an association from a particular image or image
class.  In this section we refer to {\em image n} where {\em n} is the image presented in Appendix \ref{images}.

\subsection{Descriptiveness}
We wanted to determine the likelihood of participants  assigning a textual
description to each  image type in order to measure the ease of 
{\em descriptiveness} of the image type.  Each of the 49 users in this
experiment were presented with 30 images, 6 of each image class, and
prompted to enter a description.  We gathered
1355 non-null responses (Faces: 278, Fractals: 272, Inkblots: 270,
Snowflakes: 257, Textures: 278). 
The textual descriptions assigned to image 14 give a good example of
the range of responses we 
obtained: {\em butteroad, demented frog,
mangled butterfly} and {\em angry clown}, among others. 

We found that there was a statistical difference in the number of
responses we received from users based on the image class, F(1.83,
535.4)=15.53, $p$ $<$ 0.05.  The snowflake class had a
significantly lower rate of responses when compared to all other image
classes while the face and texture classes had higher response
rates.  When we compared the response rates of individual images we
found that there were {\em no} statistical differences within the image
classes for any particular images within their class, $p$ $>$ 0.05.  This indicates that
faces and textures are the easiest image classes for users to form associations
with and that snowflakes are the most difficult.  In the following
analyses we removed all null responses and considered only responses
collected as a result of successful picture naming.

\subsection{Strength}
If we want to use the image descriptions as cues, we have to deal with
the possibility of a 
guess being made as to the image description generated by the
legitimate user. A long description, therefore, will not necessarily
act as a
strong password; one needs to consider the entropy of the description and
the variability of the responses. 

This section therefore considers the responses in terms of strength from length
(response length), image variability (character distribution and informational
entropy) and name agreement (predictability). 

\subsubsection{Response Length}
\begin{figure}[h!tbp]
\includegraphics[width=0.9\columnwidth]{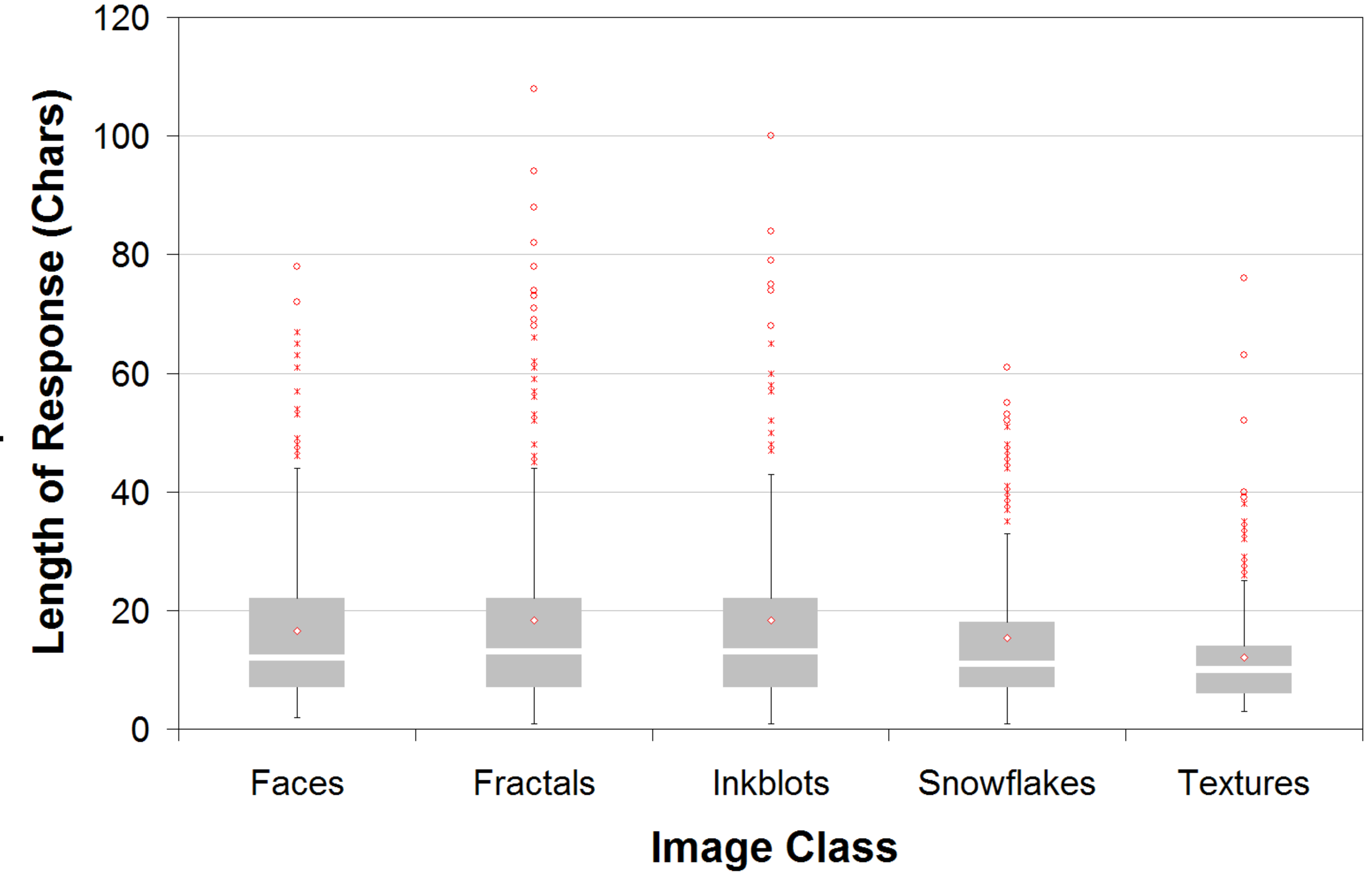} \caption[]{Response Length By Image Class}
\label{fig:length-boxplot.eps}
\end{figure}

The next measure that we consider is the average length of the textual
response obtained from an image measured in characters, in order to
determine the {\em strength} of the description if it were used as a password.
However, as we will show, since this does not take into account the
character set or probability distribution of the character set it
cannot be used to independently measure the security of a password or
textual association.  It is, nonetheless, a useful simple indicator
of security.  The results of this analysis are presented in Figure \ref{fig:length-boxplot.eps} and can be summarised as Faces (M=16.6, SE=0.83), Fractals (M=18.3, SE=1.1), Inkblots (M=18.3, SE=1.0), Snowflakes (M=15.3, SE=0.7) and Textures (M=12.1, SE=0.5)

We can observe from these results that the response
length is significantly affected by the image class, F(3.6,
1000.5)=12.414, $p$ $<$ 0.05.  The length of the response is
significantly shorter for textures than for all other image classes.    Furthermore, we found that an
extremely simple snowflake with few ``rays" had significantly lower response
character length (M=13.02, SE=1.83) than a comparable
snowflake with many rays (M=19.1, SE=1.93) indicating that overly
simple images may result in simple responses, F=(4.16,99.96)=2.85, $p$
$<$ 0.05. This shows that the type of image
shown to the user is important as different image classes can
encourage users to provide the longer responses which are generally
desired for stronger passwords.

\subsubsection{Image Variability}
{\bf Character Distribution}\\
The character distribution of the response is  a security
indicator since it gives an idea of how predictable the responses
are and how easily an image description could be guessed. 
When we considered the character distribution of the responses we
discovered that they closely parallel that of English.  This is
unsurprising as all the participants were English speakers and
since all responses would be in English, the responses would appear to
inherit the character distribution frequencies.

{\bf Informational Entropy}\\
The informational entropy of responses gives us an indication of the  image
variability of the image.  The entropy of the 
information in a signal, as defined by Shannon\cite{shannon:entropy},
specifies how much uncertainty or ``randomness" exists within the
signal.  Specifically \[ H(X) = - \sum_{i = 1}^{n} p(X_i) log_2 p(X_i)
\] where $H(X)$ is the entropy of the signal X in bits, $X_i$ is a
token in the alphabet of $X$ represented by $1..n$ and $p(X_i)$ is the
probability function representing the probability that the token will
appear in the signal.  The probability function used in this case is a
simple weight based on the character frequencies within the textual
association.  As entropy within the signal increases it becomes less
predictable and, as such, the more difficult it becomes to guess the
content.  Here we represent the entropy of a textual response by the average number of bits required to encode
each character using an encoded string.  For comparison; a standard
ASCII keyboard has 95 printable characters (including the space
character), this results in an upper bound on textual entropy of 6.57
bits per character.  The lower bound for entropy is clearly 0 bits per
character for a string composed entirely of a single character; since
the next character in the string is always predictable.  This entropic
view of textual passwords essentially measures the extent of the usage
of the available character set. 

The results of entropic analysis of the responses are summarised as
follows; 
Faces (M=3.1,SE=0.01), Fractals (M=3.1,SE=0.02), \\
Inkblots
(M=3.1,SE=0.02), 
Snowflakes (M=3,SE=0.01) \\
and Textures (M=2.9,SE=0.01).

\begin{figure}[h!tbp]
\includegraphics[width=0.9\columnwidth]{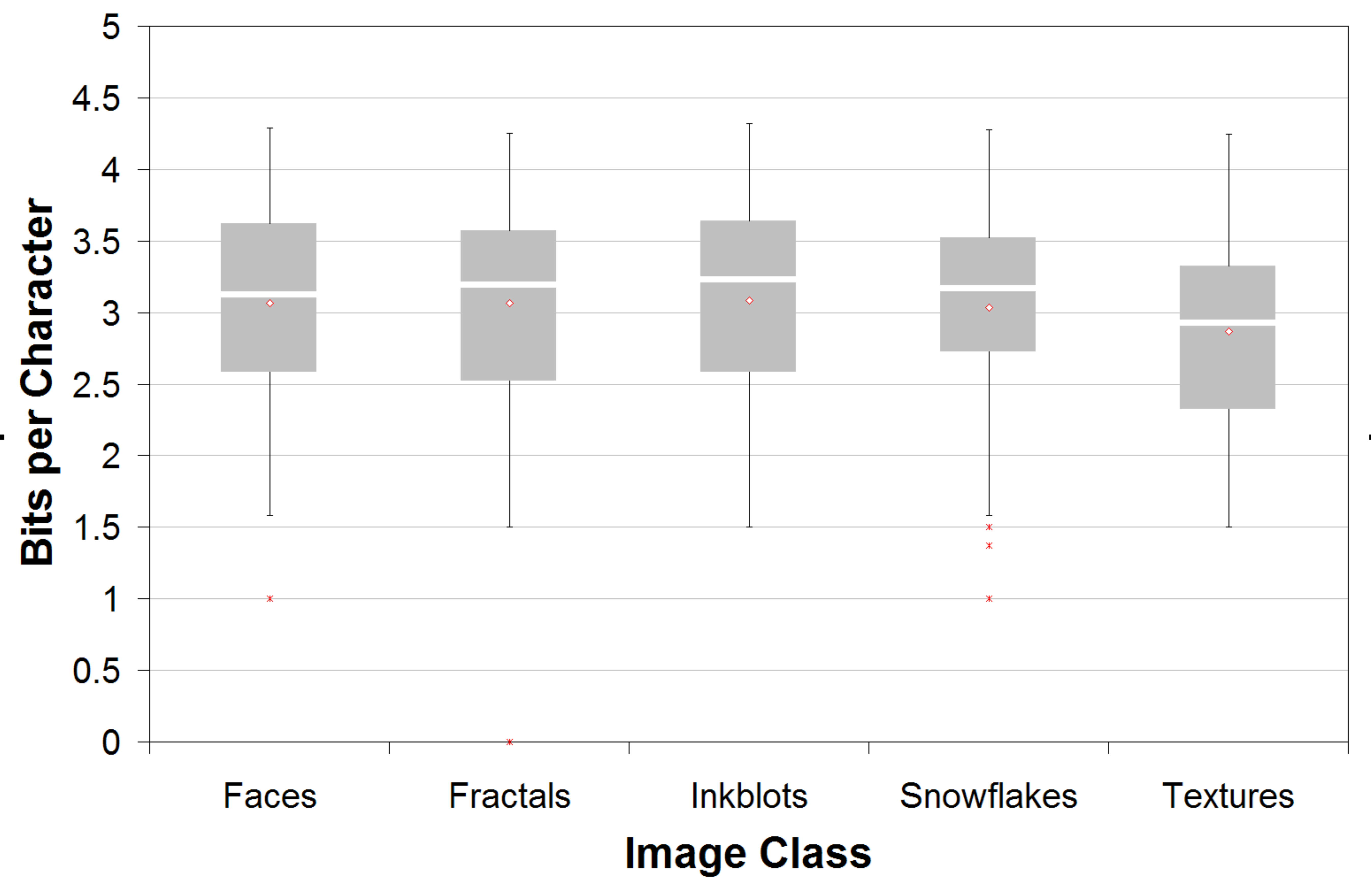} \caption[]{Bits per Character}
\label{fig:bits-per-char-boxplot.eps}
\end{figure}

From the results presented in Figure \ref{fig:bits-per-char-boxplot.eps} we can see that the texture
image class is the only class with a significant difference in the
number of bits per character, F=(4,1108)=5.49, $p$ $<$ 0.05.  We
observed that the snowflake with the highest number of rays 
had a significantly higher number of bits per character (Image 22: M = 3.28, SE
= 0.04) than three other snowflakes (Image 20: M = 2.93, SE = 0.002),(Image 21: M = 2.89,
SE = 0.05),(Image 24: M=2.93, SE=0.05), F(4.11,197.285)=4.083, $p$ $<$ 0.05.
There were significant effects between individual images within the
textures class which were caused by a single image, of a leaf (image
28), which had a significantly 
lower number of bits per character (M = 2.62, SE = 0.07).  There were
two textures with high 
amounts of repetition which had
 higher than average bits per character for the textures image
class (Image 29: M=3.00, SE=0.06),(Image 30: M=3.08, SE=0.04), F(3.753,180.12)=5.508, $p$
$<$ 0.05.

 The number of bits per character is
essentially the same for most image classes (except textures) and 
the length of the response is the largest contributor to the number of
bits per response (ie. total entropy) and therefore the overall security of a particular
textual association.

\subsubsection{Name Agreement}

We can measure name agreement using
the Smith-Waterman\cite{smith-waterman:similarity} algorithm
to measure local optimal alignments between strings.  These alignments
correspond to local similarities between strings and are a useful
measure in our case to locate instances where the strings have similar
sections --- thus measuring the similarity between two strings.  A heuristic approach was used to
determine a normalised score for each class of images normalised by response length.

The analysis shows that the Smith-Waterman score is significantly
affected by the image class, F(1.79,496)=1487, $p$ $<$ 0.05.  The
results show
that inkblots have the lowest average Smith-Water\-man score (least
similarity) followed by fractals, snowflakes, faces and finally
textures. 

Analysis of the Smith-Waterman scores for individual images 
reveals that within the faces class any
images with distinctive features (such as images 1 and 4) score higher
(more similar) Smith-Waterman scores as these features are more
readily commented upon within the user's response,
F(1.905,91.45)=36.567, $p$ $<$ 0.05. 

Two highly symmetrical fractals (images 7 and 11) had significantly
higher Smith-Water\-man scores than the other fractals,\\
F(1.725,85.62)=23.66, $p$ $<$ 0.05. 
There was a third symmetrical image within the fractal class (image
12) that did not score similarly  so the reasons for these
particularly high scores is unknown. 
There was a significant decrease in Smith-Waterman scores for the
inkblots with high density distributions of blots (images 13, 14 and
17) as compared to the more evenly distributed inkblots (images 13,
14 and 17), F(1.74,83.52)=42.196, $p$ $<$ 0.05. 
The snowflakes class exhibited significant differences in the
Smith-Waterman scores, \\
F(1.53,73.565)=66.757, $p$ $<$ 0.05.  The two
least complex  snow\-flakes (images 19 and 21) had significantly higher
(more similar) scores than all other snowflakes followed by the two
most complex images (images 20 and 24).  Interestingly the lowest
Smith-Waterman scores were for images 22 and 23 which were generated
using either the maximum number of rays or the maximum complexity but not
both. 
The images within the texture class were also found to have
significantly higher Smith-Waterman scores for easily identifiable
textures (images 28 and 29), F(1.268,60.878)=193.984, $p$ $<$ 0.05. 

In conclusion, the inkblot images scored best in terms of having a low
name agreement, followed by snowflakes while textures had the highest
level of name agreement.

\section{Discussion}\label{discuss}
When we examine the results from the previous section we discover that
the Inkblot and Fractal classes are particularly good performers for
all metrics while Texture and Snowflake classes perform poorly
(except for name agreement for the latter)

 The bits per character for each image
class was essentially the same --- indicating that response length was
the primary factor when determining the security of the image description.
Hence  for the majority of our experiments there was no
appreciable difference between {\em individual} images within the image
classes, whereas there were many differences across class boundaries.

\label{reflection}

One explanation for the above findings is that
the normal function of the visual system is to
detect strong perceptual signals, i.e. ``recognise
things'' within retinal images comprising potentially
ambiguous combinations of visual elements
set against, or even partially obscured by, background
clutter ``distractors''. The Gestalt laws indicate
grouping mechanisms that have evolved to
facilitate visual interpretation under the above conditions
and thereby improve the perceptual signal-to-noise, i.e. receiver operating characteristic, for
visual recognition. By constructing images that
potentially contain ambiguity in the arrangement
of their visual elements, we would appear to be
able to elicit ambiguity in their perceptual interpretation.

Therefore, by reducing the signal-to-noise
ratio in perceptual grouping space we can
potentially increase the range of interpretations,
and perhaps also their uniqueness as near random
groupings become associated in the mind of the
perceiver, to contribute to the security (unpredictability)
of the elicited textual responses. The
reverse of the above argument is also true; images
containing distinctive features that rise above the
perceptual signal-to-noise, limit the potential for
ambiguity and thereby multiple interpretations, as
exhibited by the texture images containing highly
distinctive elements.

On the one hand it would appear that we should
generate a combinatorial explosion of potentially
valid interpretations of the atomic visual elements
presented in each cue image. The most straightforward
approach to achieving such a combinatorial
power set would appear to be, on first sight, to
construct the cue images from a wide variety of
very small visual features, thereby maximising
their potential combinations. However, following
this approach would lead to fine texture-like fields
being generated and we observed that the text
produced in response to texture fields is not as
rich and unique as that produced by inkblots or
fractals. A potential explanation is that texture
fields are being interpreted as global homogeneous
visual percepts, e.g. classic texture fields might
elicit descriptions such as sand, water, pebbles etc,
comprising simple unitary concepts.

This observation of the effect of increasing visual complexity beyond
a certain threshold has been reported by Granovskaya {\em et
al.} \cite{Granovskaya}. Images beyond a threshold level of complexity appear to be
interpreted and memorised in terms of the statistical distribution of
atomic shapes from which they are constructed. Krienovich and Longpr\'e
\cite{kreinovich98} have suggested that limited mental memory capacity
is responsible for this barrier to memorising highly complex images
accurately and have formalised this notion in terms of a modified
version of Kolmogorov complexity. 

 By generating visual elements over a range of spatial scales, we are
 less likely to generate a homogeneous texture-like image field and
 correspondingly more likely 
to elicit local interpretations that contribute to a
global percept representing a compound object or
scene with multiple elements, to elicit the richest
responses. Hence we should maximise use of our limited mental capacity
for visual complexity by generating cue images containing structure
and structural relations between elements, as opposed to low-level
disorder. 

Fractals, by definition, are constructed from
visual elements spanning a range of spatial scales and the
inkblot generation mechanism likewise produces
and combines atomic visual elements, i.e. blots of
varying sizes. Therefore these mechanisms would
appear to be well suited to generating image fields
that contain perceptually significant local structure.
While the snowflake images also contain structures
over a range of spatial scales, their regularity and
symmetric configurations would appear to reduce
their scope for multiple competing interpretations.

The experiments in Section \ref{method} indicate that the elicited responses
are sufficiently secure to provide a viable cueing
mechanism. Given the above evidence it would appear that
at least inkblots and fractals have the potential to
serve as password cues. 
Our last metric for cue image efficacy is {\em durability}. To test
durability and hence overall efficacy in an authentication context, we used
inkblots as cues in a longitudinal experiment. 
 The following section
reports on our findings about the viability of
inkblot-like
image cues, which we have called {\em cueblots}, in eliciting strong passwords.

\section{Cueblot Authentication}\label{passwords}

In order to test the efficacy of cueblots we developed a website for an
elective module within our undergraduate computing science course. The website gave students access
to lecture notes,  their grades and various other
resources. A total of 53 undergraduate students used the website.
Users were randomly assigned to the password or cueblot conditions.
The cueblot-assisted authentication process had the following phases:

\begin{enumerate}
\item {\em Registration}: users were given a user name and
  registration code,
  by email, to facilitate the 
  registration process. When they entered the key, the system either
  allowed them to choose a password (for the password
  condition), or displayed a cueblot, and allowed the user to
  customise and tailor the cueblot, as illustrated in Figure
\ref{enrolproc}, to their satisfaction. The user was
  then instructed to give a cueblot description as a password.   The
  cueblots were comprised of 5 elements: (i) a randomly selected seed,
  (ii) the maximum diameter of blots on the canvas, (iii) the number
  of blots on the canvas, (iv) the distance between blots and (v) the
  number of colours in the cueblot.  When the user is happy with their
  choice of cueblot the system simply saves these 5 parameters which
  can be used during authentication to regenerate the cueblot.  The
  users were permitted to tailor the cueblot so  as to ensure that they were
  not presented with an cueblot that they found it impossible to create a
  textual association for.  If they were presented with a cueblot they
  considered to be obscure,
   they could either request a brand new
  one or tailor that one until they felt they were able to create an 
  association. 
\item {\em Authentication}: The users entered a user name and were the
  directed to 
  the authentication page. In the case of password users a simple
  password text entry area was supplied. In the case of cueblot users
  ``their'' cueblot was displayed
  and the user could re-enter the 
  original cueblot description.
\item {\em Replacement}: users could request a re-registration from the
  website administrator by email if the password had been forgotten.  
\end{enumerate}

\inserteps{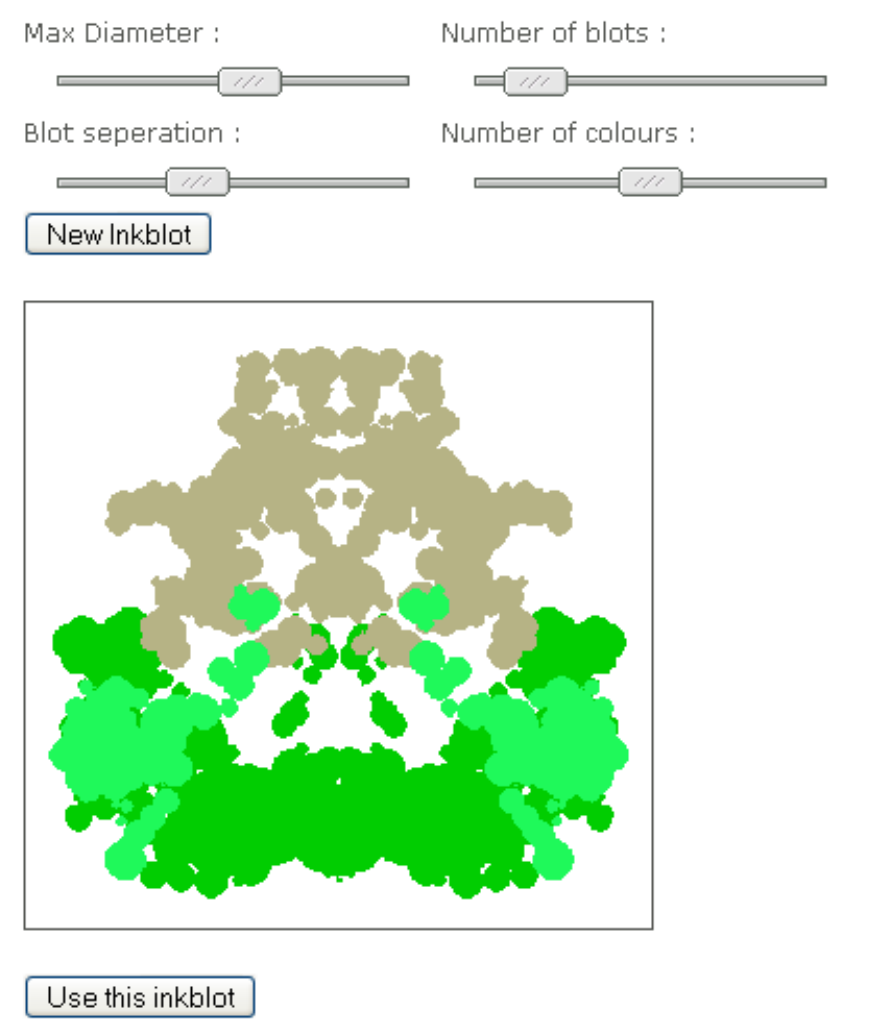}{Choosing a Cueblot at Enrolment}{enrolproc}

The experiment ran for 9 weeks and all accesses were logged to
facilitate analysis.  The results are presented in the following section. 

\subsection{Results}



 Of the users who had agreed to their
login behaviour being monitored, a total of 53 actually used the site.
Of these,  24 were allocated to the password condition and 29
to the cueblot condition.  One user from the password condition
needed a password reset during the course of the
experiment and both the original and replacement passwords are
included in our analysis; no users in the cueblot condition requested
a replacement password. 

We encountered six instances of people who deviated from the
instructions provided for their condition.  Two password users used
the registration code  as their password, probably because they had an
email record of this code, and this made things easier for them if
they forgot their password.
Four people
chose to ignore their cueblot, declining the offered cue and instead
providing a password or pass-phrase of their own choosing.  Since
  this type of behaviour is entirely possible in real life deployments, we retained
these passwords/descriptions throughout our analysis. Examples of
passwords given by cueblot users are: {\em scarypumpkin, bunnysplat, blob,
somethin} and {\em mask}. 

Authentication mechanisms, whether they make use of cues or not, must
try to maximise both security and ease of use.
The next two sections will consider our findings related to  cueblot-assisted
authentication in terms
of these
perspectives. 

\subsection{Security}

When discussing the security of an authentication scheme based on
textual input the first measure considered is typically the {\em length} of
the password and its {\em character complexity}.  That is to say; longer
passwords with larger choices of available characters (i.e. lowercase
and uppercase letters, numbers and special characters instead of just
lowercase letters) will result in much more secure passwords.  

{\bf Password Length}\\
When we
consider the length of the response for each condition we find,
surprisingly, that there is no significant difference between
passwords (M: 7.52, SE: 0.332) and cueblots (M: 8.31, SE: 0.632), $p
>$ 0.05.  Similarly when we evaluate the mean number of bits required
per character for passwords (M: 2.49, SE:
0.09) and cueblot descriptions (M: 2.64, SE: 0.11) we find that this, too, is  not
significant,  $p >$ 0.05. 

%

{\bf Password Guessability}\\
The length and number of bits per character, however, do not tell the
full story.  We also have to consider how similar descriptions are to
each other and to what extent  they have similar substrings.  

We used the Smith-Waterman
algorithm\cite{smith-waterman:similarity}, which is designed to do local sequence alignment. This
allows us to measure the longest common sequences between strings
(i.e. common uses of words such as `` the "), in this case a higher
score indicates a longer sequence and thus a {\em lower} score is desirable.
We found that cueblots (M: 0.08, SE: 0.005) had a significantly higher
Smith-Waterman score than passwords (M: 0.05, SE: 0.006), t(48.85) =
-4.088, $p$ $<$ 0.05, which indicates that users often include a larger subset of
common words within their cueblot descriptions than with traditional passwords.

%

In the next section we 
will analyse the users' performance at using cueblot-assisted authentication  in the
context of time and effort required as well as login success rates.

\subsection{Ease of Use}

In this section we focus on the results gathered from our experiment
which give us an indication of the user's experience of using the
cueblot system as
compared to the traditional password system.  

{\bf Registration}\\
Users were
 were sent a
registration code by email, which allowed them to access the site.
 Although it is often
glossed over, the registration process can play a vital role in
forming the user's initial 
perspective of the system.  In our experimental system the password
condition was a simple password entry prompt in the traditional style
(users were asked to enter the password twice to confirm it's
correctness).  By comparison, since we had elected to allow users to
design their own cueblots we implemented a cueblot designer as
part of the registration process.  We found that this resulted in users
spending considerable time designing their cueblot, inflating  the
registration time (seconds) for the cueblot condition (M: 256.03, SE:
71.364) so that it was  much more time-consuming than  password registration time (M:
44.88, SE: 8.68), t(52) = -2.729, $p < $ 0.05.  This can be viewed as
a positive or negative effect depending on the reader's point of view.
It clearly makes the registration more interactive, which is a good
thing and is likely to lead to more memorable passwords, but it does significantly
increase registration time.


{\bf Authentication}\\
We continue our discussion of time by considering the mean 
time (seconds) required to login for successful sessions.  This
measurement was taken from user name entry until 
  the login session was completed and may also include more
than one login attempt if they were unsuccessful at first.  We found
that there was a significant difference between cueblots (M: 13.08,
SE:0.532) and passwords (M: 11.15, SE:0.469), t(774) = -2.724, $p <$
0.05.  This value includes any additional time it would have taken for
the user's browser to download and display the image representing the
inkblot (generally less than 3KB in PNG format).

During the course of the experiment there were a total of 388 login
sessions for cueblots and 412 login sessions for passwords.  Of these
there were a significantly lower number of sessions with a login
failure for the cueblot condition (23) than for the password condition
(44), $p <$ 0.05.  This puts the mean number of failed sessions for
passwords at 11\% and cueblots at 6\%.

%
%
%
%

However, when we look at the failed sessions in more detail we
discover that within a login attempt session the average number of
attempts at getting the  password correct per session is somewhat
different.  We found 
that there was an average of 1.18 attempts (SE: 0.118) per session for
a password while cueblots required an average of 1.96 attempts (SE:
0.493).  Our results  indicate that there was borderline
significance (t(65) = -1.985, $p =$ 0.051) which may warrant further
investigation.  Thus cueblot users are more likely to get it
right first time but may make more attempts to login if they fail the
first time. 

We also considered the number of sessions which were regarded as
``total failures" ie. sessions within which there was a failed login
attempt (or a
sequence of failed login attempts) but no eventual success indicating
that the user gave up.  We found that there was no significant difference
in this respect (3 failed cueblot sessions, 2 failed password
sessions, $p >$ 0.05). 


%
%

\subsection{Are Cueblots Efficacious Password Cues?}
Efficacy metrics, as outlined in Section \ref{imagcueing}, are {\em descriptiveness,
strength} and {\em durability}. 
In terms of descriptiveness and strength, these results appear to
conflict with the results of 
our previous 
experiment.  The length of response decreased significantly once
users were asked to perform this task within a live authentication
system, and this impacts the strength of the password.  
Furthermore, provided textual descriptions were shorter and less descriptive,
and, indeed, some appeared to have nothing to do with the provided
cue, so the cueblot fails the descriptiveness test  as well. 
This result strengthens findings by Brostoff {\em et al.}
\cite{brostoff:passfacesusability} 
during evaluation of the
Passfaces{\footnote{\url{http://www.realuser.com}}} authentication
mechanism where usage of an authentication system in real life differed
significantly from lab-based experimentation. 

In {\em our} experiment 
it seems that when the user knows that the cueblot description is going to be
used frequently as a password, he or she
 provides a much shorter description than would be provided if the
 description was only going to be provided once or twice in a
 lab-based experiment. This is perfectly
 reasonable, because users emphasise convenience over security. 
Hence the  length of
response and bits per char are basically the same as passwords. This
is rather disappointing since we had hoped that the presence of the
cueblot would allay users' fears of forgetting their passwords and
therefore encourage them to choose longer (and stronger) passwords.  

Finally, as regards durability, the cueblot users {\em did} appear to
have less trouble remembering their textual descriptions, although
this obviously does not apply to the four users who did not provide a
cueblot description, but rather provided a non-cued password.

When we first began our research into this area we believed that
users would embrace the ability to create longer passwords if they
were provided with a way to help them remember the password more easily.
Unfortunately this does not appear to be the case, confirming the
findings of Dhamija and Perrig \cite{Dhamija00} that people are only
willing to expend the minimum effort in managing their passwords.  Our results 
 indicate that the descriptions offered by users
 are of comparable length and complexity to
traditional passwords but with the problem that they will tend to
include common stop-words in their description which weakens the
password.  

The general trend of using sub-optimal passwords accords well with
Payne's \cite{Payne82} findings about how people conduct an implicit
cost-benefit analysis when making decisions and choices. Users of
passwords are clearly trading off the extra effort involved in typing
in long and complicated passwords as against the risk of having an
intruder break into their account. The risk is obviously not high
enough for them to put the extra effort in, to use longer and
stronger passwords. 

\section{Conclusion}\label{conc}
We investigated the hypothesis that  passwords could be cued by using
suitably chosen images. In Section \ref{results} we
found that 
 textual descriptions elicited by some of the images types did indeed show some
 potential.
In particular, we found that inkblot-type abstract images elicited the
longest and strongest textual descriptions. We therefore decided to
conduct a further experiment in order to test the durability of these
inkblot-type images. 

However, in our final experiment,
 we found  that the presence of the cueblot did not have a positive
 effect on the users. It did not appear to encourage them to 
 strengthen their passwords and they did not  exploit the true potential of their
 cueblot in coming up with a textual description thereof, probably
 because users anticipate the extra effort involved in continuously
 entering the 
 long description at each authentication attempt with 
 little enthusiasm.

We have to conclude that, whereas the cueblots were theoretically
viable in terms of cueing passwords, the end-user's desire for
convenience and speed of access led them not to exploit the potential
for cueing provided by the cueblot. Perhaps the only conclusion is
that the combination of convenience-seeking users and passwords is
doomed to failure. If this is the case then any auxiliary efforts to
strengthen the mechanism, such as the one explored in this paper, are
futile. 

However, there is are other contexts within which cueblots could well
be efficacious. For example,  cueblots could be used instead of the security
questions that are currently used when users forget their
passwords. Their descriptions are held only by the system itself, and
therefore could not be discovered by judicious use of the person's
social networking page or personal website \cite{Rabkin08}. Further experiments will focus
on the use of cueblots in other contexts, since we believe that they
do offer much potential due to their superior descriptiveness and
strength. 

\appendix

\section{Images}
\label{images}

\subsection{Faces}

The face images used in this study were collected from the Essex
University Computer Vision Facial Databases\cite{essex:facedatabase}
``face94" and ``face95" and were chosen to represent an equal mix of
male and female faces with a range of physical features.  Only images
that were clearly visible with similar scale and without distracting
backgrounds were considered. 

\begin{figure}[H]
\begin{minipage}[t]{2cm}
\includegraphics[width=0.9\textwidth,height=0.9\textwidth]{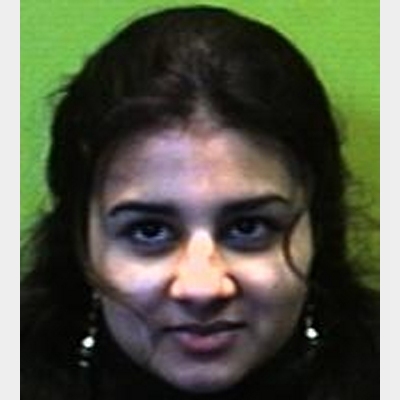} \caption[]{Face 1 (Image 1)} \label{fig:face001.eps}
\end{minipage}
\hfill
\begin{minipage}[t]{2cm}
\includegraphics[width=0.9\textwidth,height=0.9\textwidth]{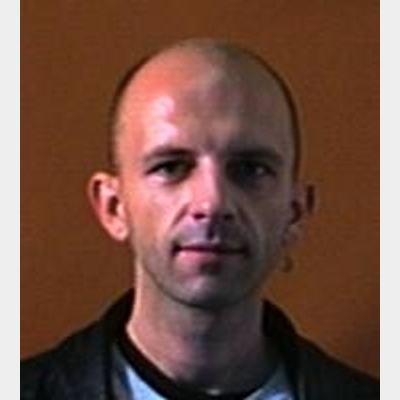} \caption[]{Face 2 (Image 2)} \label{fig:face002.eps}
\end{minipage}
\hfill
\begin{minipage}[t]{2cm}
\includegraphics[width=0.9\textwidth,height=0.9\textwidth]{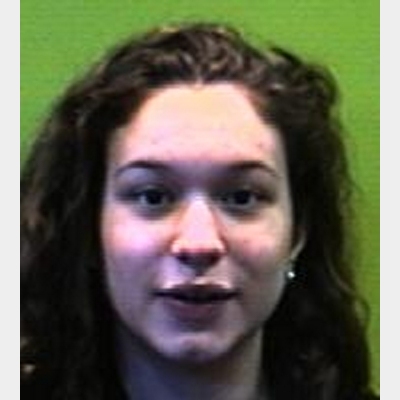} \caption[]{Face 3 (Image 3)} \label{fig:face003.eps}
\end{minipage}
\hfill
\begin{minipage}[t]{2cm}
\includegraphics[width=0.9\textwidth,height=0.9\textwidth]{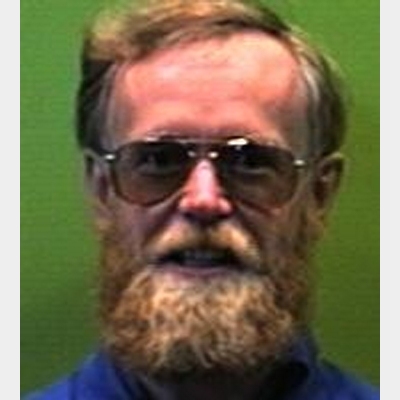} \caption[]{Face 4 (Image 4)} \label{fig:face004.eps}
\end{minipage}
\hfill
\vspace{0.25cm}
\begin{minipage}[t]{2cm}
\hfill
\end{minipage}
\hfill
\begin{minipage}[t]{2cm}
\includegraphics[width=0.9\textwidth,height=0.9\textwidth]{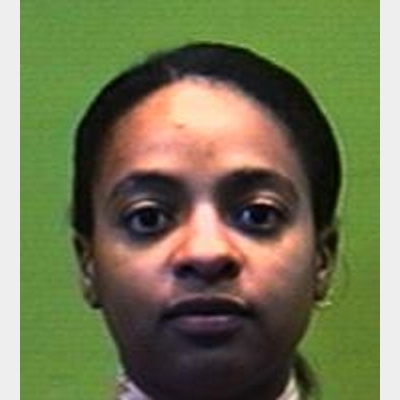} \caption[]{Face 5 (Image 5)} \label{fig:face005.eps}
\end{minipage}
\hfill
\begin{minipage}[t]{2cm}
\includegraphics[width=0.9\textwidth,height=0.9\textwidth]{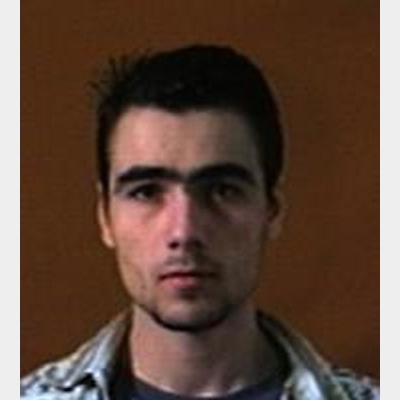} \caption[]{Face 6 (Image 6)} \label{fig:face006.eps}
\end{minipage}
\hfill
\begin{minipage}[t]{2cm}
\hfill
\end{minipage}
\hfill
\end{figure}

\subsection{Fractal}

The Fractals were generated using a commercial program Ultra Fractal\cite{slijkerman:ultrafractal}.  Variations within the image class were obtained by changing the algorithm used to generate the fractal in addition to varying the viewing position and colouring algorithms.

\begin{figure}[H]
\begin{minipage}[t]{2cm}
\includegraphics[width=0.9\textwidth,height=0.9\textwidth]{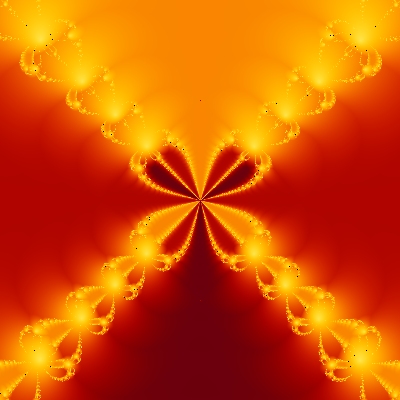} \caption[]{Fractal 1 (Image 7)} \label{fig:fractal001.eps}
\end{minipage}
\hfill
\begin{minipage}[t]{2cm}
\includegraphics[width=0.9\textwidth,height=0.9\textwidth]{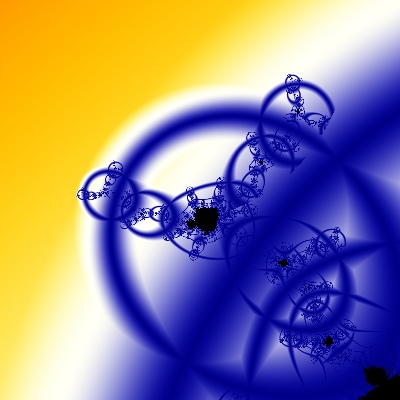} \caption[]{Fractal 2 (Image 8)} \label{fig:fractal002.eps}
\end{minipage}
\hfill
\begin{minipage}[t]{2cm}
\includegraphics[width=0.9\textwidth,height=0.9\textwidth]{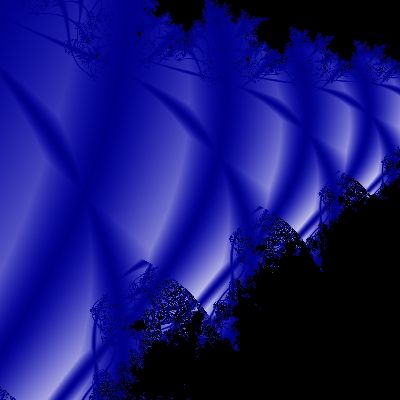} \caption[]{Fractal 3 (Image 9)} \label{fig:fractal003.eps}
\end{minipage}
\hfill
\begin{minipage}[t]{2cm}
\includegraphics[width=0.9\textwidth,height=0.9\textwidth]{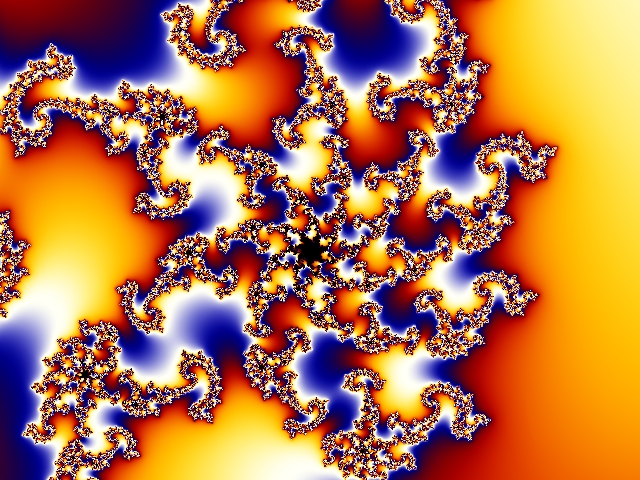} \caption[]{Fractal 4 (Image 10)} \label{fig:fractal004.eps}
\end{minipage}
\hfill
\vspace{0.25cm}
\begin{minipage}[t]{2cm}
\hfill
\end{minipage}
\hfill
\begin{minipage}[t]{2cm}
\includegraphics[width=0.9\textwidth,height=0.9\textwidth]{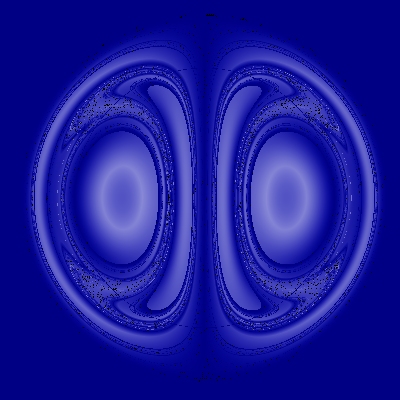} \caption[]{Fractal 5 (Image 11)} \label{fig:fractal005.eps}
\end{minipage}
\hfill
\begin{minipage}[t]{2cm}
\includegraphics[width=0.9\textwidth,height=0.9\textwidth]{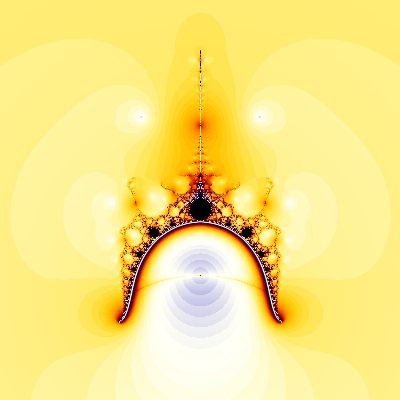} \caption[]{Fractal 6 (Image 12)} \label{fig:fractal006.eps}
\end{minipage}
\hfill
\begin{minipage}[t]{2cm}
\hfill
\end{minipage}
\hfill
\end{figure}

\subsection{Inkblots}

The inkblots were generated by a custom PHP script.  The inkblots were built by dropping ``blots" onto a canvas and ensuring the next blot landed within a fixed area of the previous blot.  The canvas was then mirrored to create the final inkblot.  The images were varied by changing the values of variables which control the number of blots, blot diameter, colour and distance between the blots.

\begin{figure}[H]
\begin{minipage}[t]{2cm}
\includegraphics[width=0.9\textwidth,height=0.9\textwidth]{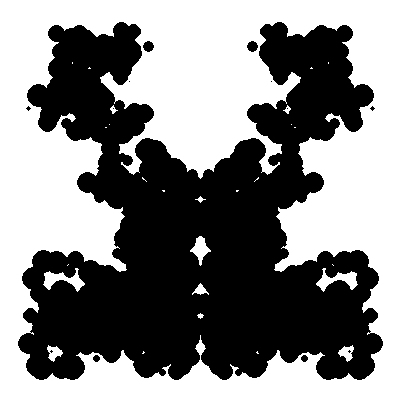} \caption[]{Inkblot 1 (Image 13)} \label{fig:inkblot001.eps}
\end{minipage}
\hfill
\begin{minipage}[t]{2cm}
\includegraphics[width=0.9\textwidth,height=0.9\textwidth]{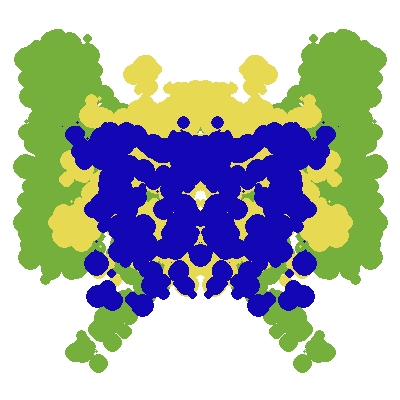} \caption[]{Inkblot 2 (Image 14)} \label{fig:inkblot002.eps}
\end{minipage}
\hfill
\begin{minipage}[t]{2cm}
\includegraphics[width=0.9\textwidth,height=0.9\textwidth]{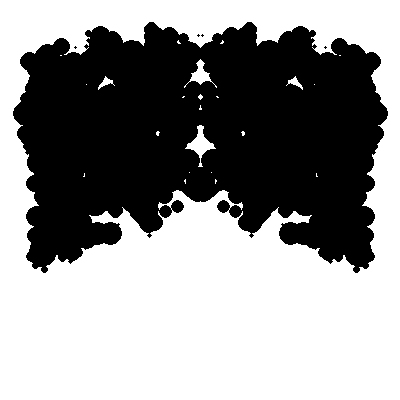} \caption[]{Inkblot 3 (Image 15)} \label{fig:inkblot003.eps}
\end{minipage}
\hfill
\begin{minipage}[t]{2cm}
\includegraphics[width=0.9\textwidth,height=0.9\textwidth]{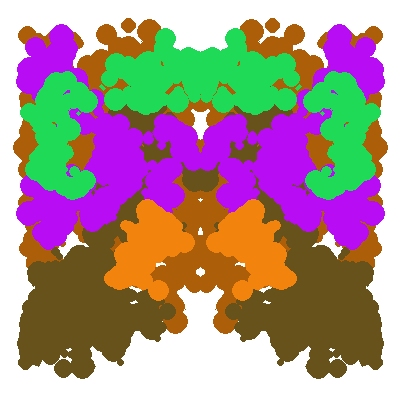} \caption[]{Inkblot 4 (Image 16)} \label{fig:inkblot004.eps}
\end{minipage}
\hfill
\vspace{0.25cm}
\begin{minipage}[t]{2cm}
\hfill
\end{minipage}
\hfill
\begin{minipage}[t]{2cm}
\includegraphics[width=0.9\textwidth,height=0.9\textwidth]{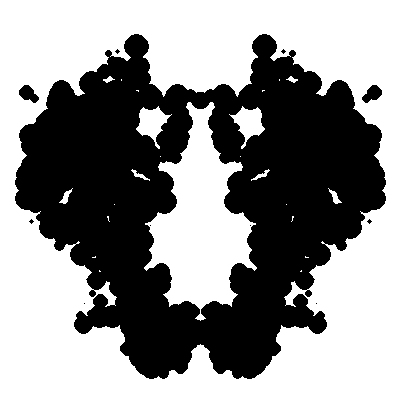} \caption[]{Inkblot 5 (Image 17)} \label{fig:inkblot005.eps}
\end{minipage}
\hfill
\begin{minipage}[t]{2cm}
\includegraphics[width=0.9\textwidth,height=0.9\textwidth]{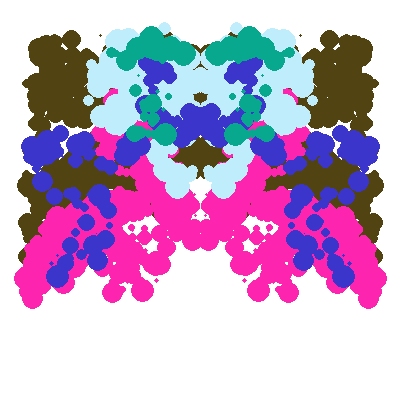} \caption[]{Inkblot 6 (Image 18)} \label{fig:inkblot006.eps}
\end{minipage}
\hfill
\begin{minipage}[t]{2cm}
\hfill
\end{minipage}
\hfill
\end{figure}

\subsection{Snowflakes}

The Snowflake images were generated using A.I. Studio Snowflake Generator\cite{aistudio:snowflakegen} and variations within the images were achieved primarily by varying the number and complexity of the rays along with scaling and position details.

\begin{figure}[H]
\begin{minipage}[t]{2cm}
\includegraphics[width=0.9\textwidth,height=0.9\textwidth]{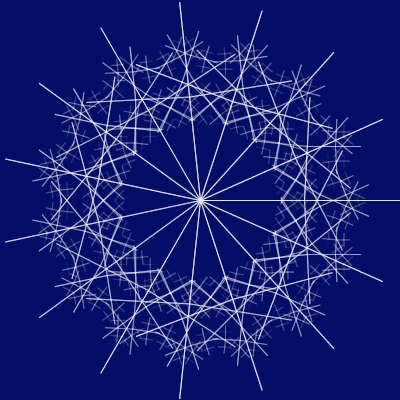} \caption[]{Snowflake 1 (Image 19)} \label{fig:snowflake001.eps}
\end{minipage}
\hfill
\begin{minipage}[t]{2cm}
\includegraphics[width=0.9\textwidth,height=0.9\textwidth]{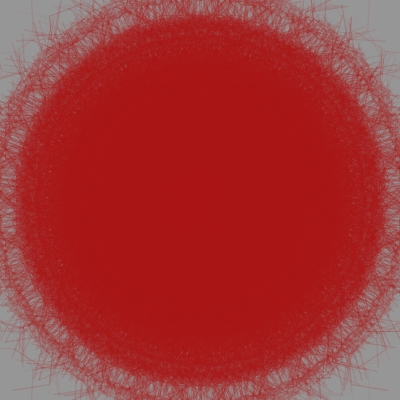} \caption[]{Snowflake 2 (Image 20)} \label{fig:snowflake002.eps}
\end{minipage}
\hfill
\begin{minipage}[t]{2cm}
\includegraphics[width=0.9\textwidth,height=0.9\textwidth]{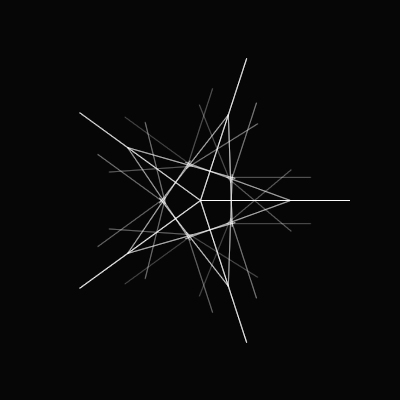} \caption[]{Snowflake 3 (Image 21)} \label{fig:snowflake003.eps}
\end{minipage}
\hfill
\begin{minipage}[t]{2cm}
\includegraphics[width=0.9\textwidth,height=0.9\textwidth]{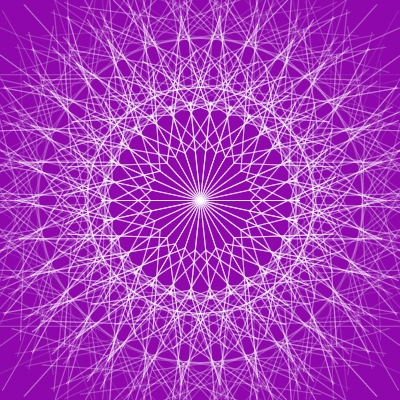} \caption[]{Snowflake 4 (Image 22)} \label{fig:snowflake004.eps}
\end{minipage}
\hfill
\vspace{0.25cm}
\begin{minipage}[t]{2cm}
\hfill
\end{minipage}
\hfill
\begin{minipage}[t]{2cm}
\includegraphics[width=0.9\textwidth,height=0.9\textwidth]{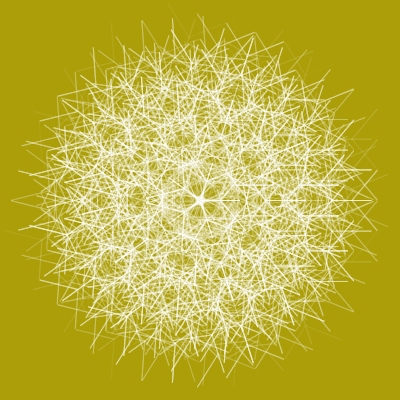} \caption[]{Snowflake 5 (Image 23)} \label{fig:snowflake005.eps}
\end{minipage}
\hfill
\begin{minipage}[t]{2cm}
\includegraphics[width=0.9\textwidth,height=0.9\textwidth]{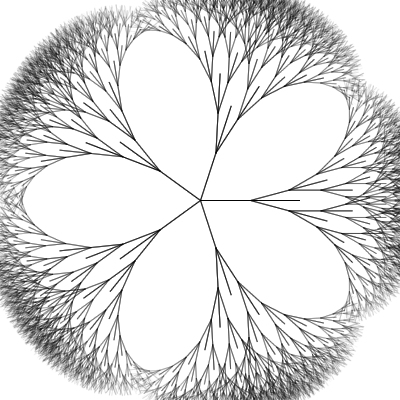} \caption[]{Snowflake 6 (Image 24)} \label{fig:snowflake006.eps}
\end{minipage}
\hfill
\begin{minipage}[t]{2cm}
\hfill
\end{minipage}
\hfill
\end{figure}

\subsection{Textures}

The Texture images were obtained from the CUReT\cite{curet} texture database and were chosen to represent a range of different textures including both man-made and natural textures.

\begin{figure}[H]
\begin{minipage}[t]{2cm}
\includegraphics[width=0.9\textwidth,height=0.9\textwidth]{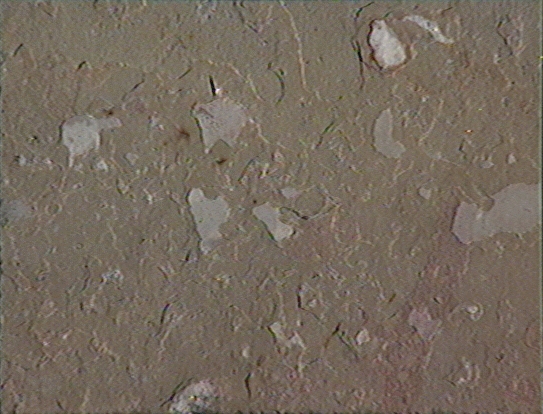} \caption[]{Texture 1 (Image 25)} \label{fig:texture001.eps}
\end{minipage}
\hfill
\begin{minipage}[t]{2cm}
\includegraphics[width=0.9\textwidth,height=0.9\textwidth]{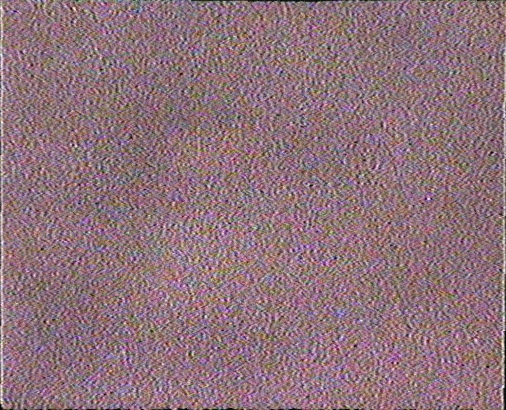} \caption[]{Texture 2 (Image 26)} \label{fig:texture002.eps}
\end{minipage}
\hfill
\begin{minipage}[t]{2cm}
\includegraphics[width=0.9\textwidth,height=0.9\textwidth]{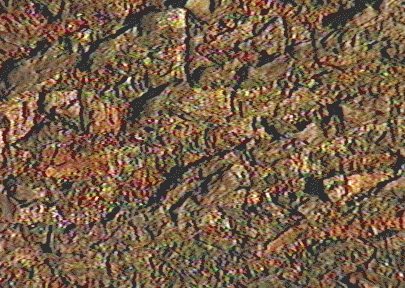} \caption[]{Texture 3 (Image 27)} \label{fig:texture003.eps}
\end{minipage}
\hfill
\begin{minipage}[t]{2cm}
\includegraphics[width=0.9\textwidth,height=0.9\textwidth]{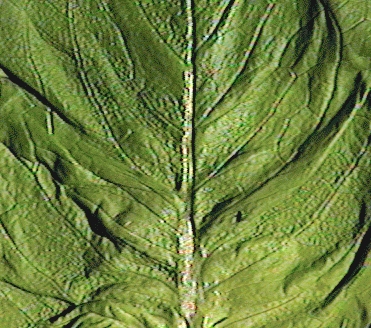} \caption[]{Texture 4 (Image 28)} \label{fig:texture004.eps}
\end{minipage}
\hfill
\vspace{0.25cm}
\begin{minipage}[t]{2cm}
\hfill
\end{minipage}
\hfill
\begin{minipage}[t]{2cm}
\includegraphics[width=0.9\textwidth,height=0.9\textwidth]{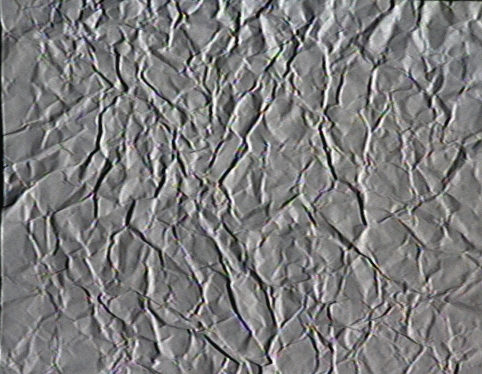} \caption[]{Texture 5 (Image 29)} \label{fig:texture005.eps}
\end{minipage}
\hfill
\begin{minipage}[t]{2cm}
\includegraphics[width=0.9\textwidth,height=0.9\textwidth]{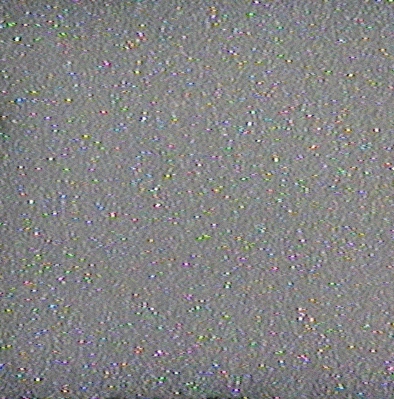} \caption[]{Texture 6 (Image 30)} \label{fig:texture006.eps}
\end{minipage}
\hfill
\begin{minipage}[t]{2cm}
\hfill
\end{minipage}
\hfill
\end{figure}

\bibliographystyle{elsart-num}
\bibliography{paper}

\end{document}